# Ancestry-specific analyses of genome-wide data confirm the settlement sequence of Polynesia

Alexander G. Ioannidis[1*], Javier Blanco-Portillo[2], Erika Hagelberg[3], Juan Esteban Rodríguez-Rodríguez[2], Keolu Fox[4], Adrian V. S. Hill[5,6], Carlos D. Bustamante[1], Marcus W. Feldman[2], Alexander J. Mentzer[5], Andrés Moreno-Estrada[7*]

[1]Department of Biomedical Data Science, Stanford Medical School, Stanford, CA, USA
[2]Department of Biology, Stanford University, Stanford, CA, USA
[3]Department of Biosciences, University of Oslo, Oslo, Norway
[4]Department of Anthropology, University of California San Diego, La Jolla, CA, USA
[5]Wellcome Centre for Human Genetics, University of Oxford, Roosevelt Drive, Oxford, UK
[6]The Jenner Institute, Nuffield Department of Medicine, University of Oxford, Oxford, UK
[7]National Laboratory of Genomics for Biodiversity (LANGEBIO), CINVESTAV, Irapuato, Guanajuato, Mexico
*Corresponding authors – Email: ioannidis@stanford.edu and andres.moreno@cinvestav.mx

Abstract: By demonstrating the role that historical population replacements and waves of admixture have played around the world, the prior work of Reich and colleagues has provided a paradigm for understanding human history [Reich et al. 2009; Reich et al. 2012; Patterson et al. 2012]. Although we show in Ioannidis et al. [2021] that the peopling of Polynesia was a range expansion, and not, as suggested by Huang et al. [2022], yet another example of waves of admixture and large-scale gene flow between populations, we believe that our result in this recently settled oceanic expanse is the exception that proves the rule.

In Ioannidis et al. [2021], we used genotypes from 430 modern individuals, allele frequency statistics, and the Chu-Liu-Edmonds algorithm [Chu & Lui, 1965; Edmonds, 1967] to reconstruct the branching settlement path that peopled the Polynesian islands. We demonstrated strong bottleneck effects, the sign of a serial founder process, by analyzing the increase in frequency of rare variants, genetic surfing, and runs of homozygosity. In our main figure (Figure 2 of Ioannidis et al. [2021]), we used the directionality statistic $\psi$ of Peter et al. [Peter & Slatkin, 2013; Peter & Slatkin, 2015] to show that there was a range expansion originating in the western islands of Samoa and Tonga that resulted in the peopling of the remote eastern Polynesian islands. We discussed the continued cultural interaction and influences between some islands and cited archaeological analyses of stone tools traded between those islands, but noted it was unlikely that voyages passing across thousands of kilometers of open ocean could carry sufficient people to have had a significant genetic impact on post-expansion island populations' allele frequencies. Crews, however, could, and did, have a significant cultural impact (see our Supplementary Information section titled "On Cultural Clusters and Diffusion Versus Genetic Clusters") and left some impact on identity-by-descent (IBD) fragments, which we discussed. As we stated, "Although the settlement path followed by the Polynesian genetic component found on each island is, in the absence of population replacement, the founding migration path, this settlement path is not necessarily coincident with the various later inter-island cultural diffusion paths (of language, cosmogony, and customs)."

Huang et al. [2022] contest our settlement path conclusions without discussing any of our path reconstruction methods above (Chu-Liu-Edmonds algorithm, principal curves, and tree-based methods), each based on allele frequencies, and our genetic bottleneck and isolation findings, the two main thrusts of our paper. Huang et al. [2022] refer to inter-island trade interactions, such as the trade in stone tools that we cited [Rolett, 2002; Collerson & Weisler, 2007; Weisler et al. 2016] and assert that we do not test for a model of inter-island gene flow postdating settlement: "One process not modelled by Ioannidis et al. [2021] is gene flow postdating the initial settlements of Polynesia." We agree that such a model is important to test, and at a reviewer's request (see online open reviews), we did so. As we described in the Online Methods section of our 2021 paper, we used the F4-statistic [Patterson et al. 2012; Peter, 2016] to test for potential gene flow between islands postdating the initial settlement of Polynesia and found that (except for a few exceptions, which we discuss) there was no detectable signal on allele frequencies of significant post-settlement gene flow between islands. This test contradicts the conclusions of Huang et al [2022].

Huang et al. [2022] focus on an unrelated analysis (based on IBD fragments) that we perform later in the text and which constitutes a subsidiary part of our paper. We used this analysis to add date annotations to the migration paths, whose routes we had already derived using the separate allele frequency methods above. As we stated, "Note that our reconstruction of the settlement path is independent of these date estimates, which are overlaid upon it, and is more robust to later sporadic contact than IBD distributions are (see Methods sections 'Polynesian ancestry-specific allele frequency analyses' and 'F4')."

We further cautioned that our dates were only terminal bounds to the settlement date of each island, referring to them as "terminus ante quem" in our text. Huang et. al [2022] suggest that looser (later) bounds for these dates are called for and derive a population-size-dependent estimator for these looser upper bounds. This derivation is valuable. However, ultimately Huang et al. [2022] do not believe that their looser (later) bounds represent the actual island settlement dates, and they repeat our original hypothesis that such IBD-based dates can be influenced by later inter-island trade. Thus, although Huang et al. [2022] relax our bounds on settlement dates, they appear to agree with us that the islands were settled earlier than their IBD-based dates.

Huang et al. [2022] also differ with us in their hypothesis concerning the origin of island megaliths. Our hypothesis that megaliths could be associated with a common founding population was based on our observation that megaliths are located only on islands found on a single branch of our allele-frequency-based settlement path reconstruction. Huang et al. [2022] do not discuss our settlement path reconstruction, which is the main body of evidence in our paper and the basis for our hypothesis about the megaliths. Nevertheless, they assert a competing speculation that the presence of megaliths across the particular islands that we identified stems from later inter-island cultural interactions. Even without genetic interactions, later inter-island cultural interactions could have spread these megaliths, as we pointed out in our Supplementary Information section titled "On Cultural Clusters and Diffusion Versus Genetic Clusters." Trade interactions between many of the islands have already been archaeologically confirmed, as we discussed in our paper [Rolett, 2002; Collerson & Weisler, 2007; Weisler et al. 2016], so it is unclear how the IBD date arguments of Huang et al. [2022] would answer this cultural debate. In any case, both competing hypotheses about the megaliths remain to be tested. Our hypothesis is supported by a correlation of the megaliths with our settlement path reconstruction, while the hypothesis of Huang et al., as they currently state it, has no particular supporting evidence of associations.

Other concerns exist. Based on their dating analyses, Huang et. al [2022] conclude that significant genetic interactions continued between Easter Island and other islands well into the 16th century. This is inconsistent with the Rapanui language's linguistic divergence [Kirch, 2017], our F4-statistic allele frequency test (omitted from the discussion of Huang et al. [2022] and contradicting their thesis of significant inter-island genetic interaction), our analysis of ancient Rapanui DNA (demonstrating genetic continuity on the island across time, rather than genetic shifts from gene flow), and the main results of our paper; namely, that there is a clear decrease in genetic diversity on Easter Island [Ioannidis et al. 2021]. Indeed, the range expansion statistic ψ (based on allele frequencies and independent of the IBD analyses that Huang et al. discuss) shows a clear directionality of settlement (Figure 2 of Ioannidis et al. [2021]), rather than a network of significant, population-wide genetic intermingling between islands.

Our approach to benchmark our settlement dates was empirical, confirmed with known dates for the settlement of Easter Island (based on C-14 dating and pollen dating [Hunt & Lipo, 2008; Mulrooney, 2013], see also Figure 1 below). That island is believed to have been too remote to have had significant continued interactions with other Polynesian islands [Barnes, 2006; Hunt & Lipo, 2011; Weisler et al. 2016; Kirch, 2017; Cochrane & Hunt, 2018]. This allowed us to use Easter Island (settled c. 1200 AD) as a benchmark grounding our method and then to use IBD length decay distributions to infer which islands were settled earlier or later and by how much (Extended Data Table 1 of Ioannidis et al. [2021]), relying on a logarithmic scaling that is to first order the same as that of the formula of Huang et al. [2022] when expanded around this date. Estimating settlement times using the approach of Huang et al. [2022] also requires a benchmark date in order to estimate continuous migration rates (as they show in their Figure S4) or colonial era migration pulses (which they do not model), so their method does not seem to offer a notable advantage for empirical settlement date estimation in this dataset in practice.

Finally, our truncation of the exponential fit at 15 cM, mentioned by Huang et al. [2022], was performed because segments >15 cM are likely post-European-contact and such long outliers could particularly bias the model. After the arrival of Europeans, it is known there was larger-scale inter-island transport and colonial mediated interaction throughout Polynesia [Kirch, 2017].

In conclusion, as we stated in Ioannidis et al. [2021] and as Huang et al. [2022] concur, IBD-based dating in this study can only be taken as a terminus ante quem due to the existence of some known inter-island interactions, but as we further stated in our 2021 paper, "It is important to note that these date estimates are not used at any stage for finding the sequence of islands settled, which is instead derived from the genetic-based directionality statistic (ψ) that measures the telescoping series of founding bottlenecks. Thus, any variation in dates due to the above sources of uncertainty will not affect the inferred settlement sequence." Our assessment of island settlement paths, and our consequent hypotheses about island megaliths, are based on allele frequency variation, not on IBD, are supported by a negative test for significant influence of inter-island gene flow on allele frequencies using the F4-statistic, and are confirmed with comparison of modern islander allele frequencies to ancient DNA samples from those islands (Ioannidis et al. [2021], Extended Data Figure 3 "Continuity between ancient and modern Polynesian island populations"). The IBD derivation of Huang et al. [2022] is valuable, but the authors concede that their equation is not suited to directly determining the island settlement dates, and it does not resolve different demographic scenarios (e.g. the continuous migration model they advance versus more recent, known colonial-era interactions that we discuss). We do not see how it invalidates our conclusions for this dataset, which are based on more robust allele frequency methods reinforced by a direct comparison to ancient DNA.

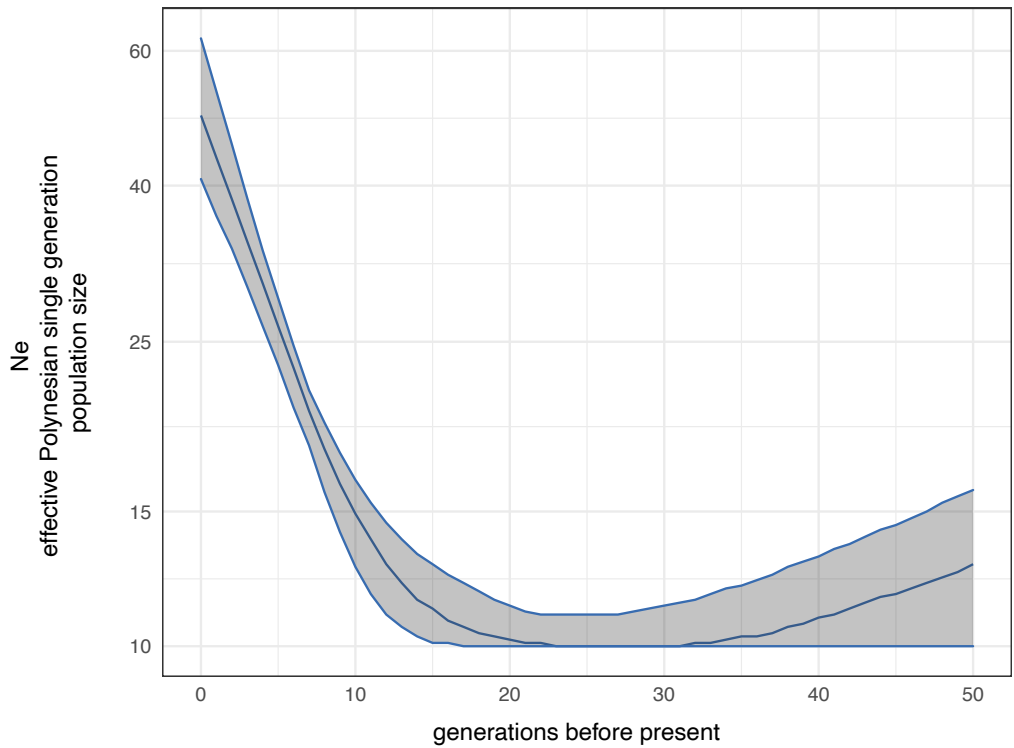

Figure 1: **Rapa Nui (Easter Island) single generation effective population size inferred from Polynesian IBD segments.** The central curve illustrates the effective historical Polynesian population size on the island of Rapa Nui (Easter Island) inferred from the shared IBD ancestry segments between 166 modern Rapanui samples analyzed in Ioannidis et al. [2021] using the ancestry specific IBDNe method of Browning et al. [2018]. The shaded region represents the 95% confidence interval. To compute the actual population size one must account for the fact that multiple generations are alive at any one time (children, parents, and grandparents) and multiply by about three [Browning & Browning, 2015]. (To compute the present-day census size on Rapa Nui one would also have to further adjust upwards, dividing by the island's average Polynesian ancestry fraction [Browning et al. 2018], as the islanders are no longer only of Polynesian ancestry.) A clear founding bottleneck on Rapa Nui is seen, achieving a minimum at 27 generations ago. This corresponds to ~1200 AD, using thirty years per generation (parental age averaged across fathers and mothers and all lifetime children [Matsumura, 2008; Helgason, 2003; Fenner, 2005]) and the sample collection date of 2013. This result supports the strong bottleneck effect characterized in Ioannidis et al. [2021], agrees with archeological date estimates for Rapa Nui settlement from pollen and C-14 dating [Hunt & Lipo, 2008; Mulrooney, 2013], and confirms the settlement date benchmark used in Ioannidis et al. [2021]. IBD segments were called for segments >3 cM using the new, more accurate HapIBD methodology of Zhou et al. [2020] with default parameters (min-seed=2.0, min-extend=1.0,2.0, min-mac=2.0, min-markers=100, and max-gap=1000) and were then filtered for only genomic segments of Polynesian ancestry (excluding European and Native American ancestry segments), as described in Ioannidis et al. [2021]. Fragments were merged when the length of gap between IBD segments was <=0.2 cM if the IBD gap had a maximum number of one genotype inconsistent with IBD. IBD segments located in centromeres and telomeres were removed, resulting in a final 505,270 total segments for analysis. IBDNe was run with parameters: mincm=3.0, minregion=50.0, trimcm=0.2, filtersample=false, npairs=1565, nits=1000, nboots=80, and gmax=200.